

\def\oneandahalfspace{\baselineskip=\normalbaselineskip
  \multiply\baselineskip by 3 \divide\baselineskip by 2}

\parskip=\medskipamount
\overfullrule=0pt
\raggedbottom
\def\normalparindent{24pt}
\nopagenumbers
\footline={\ifnum\pageno=1{\hfil}\else{\hfil\rm\folio\hfil}\fi}
\def\endpage{\vfill\eject}
\def\beginlinemode{\endmode\begingroup\parskip=0pt
                   \obeylines\def\\{\par}\def\endmode{\par\endgroup}}
\def\beginparmode{\endmode\begingroup \def\endmode{\par\endgroup}}
\let\endmode=\par
\def\raggedcenter{
                  \leftskip=2em plus 6em \rightskip=\leftskip
                  \parindent=0pt \parfillskip=0pt \spaceskip=.3333em
                  \xspaceskip=.5em\pretolerance=9999 \tolerance=9999
                  \hyphenpenalty=9999 \exhyphenpenalty=9999 }
\def\\{\cr}
\let\rawfootnote=\footnote\def\footnote#1#2{{\parindent=0pt\parskip=0pt
        \rawfootnote{#1}{#2\hfill\vrule height 0pt depth 6pt width 0pt}}}
\def\title{\null\vskip 3pt plus 0.2fill\beginlinemode\raggedcenter\bf}
\def\author{\vskip 3pt plus 0.2fill \beginlinemode\raggedcenter}
\def\affil{\vskip 3pt plus 0.1fill\beginlinemode\raggedcenter\it}
\def\abstract{\vskip 3pt plus 0.3fill \beginparmode{\noindent  ABSTRACT:~}  }
\def\endtitlepage{\endpage\body}
\def\body{\beginparmode\parindent=\normalparindent}
\def\head#1{\par\goodbreak{\immediate\write16{#1}
           {\noindent\bf #1}\par}\nobreak\nobreak}

\def\refto#1{$^{[#1]}$}
\def\ref#1{Ref.~#1}
\def\Ref#1{Ref.~#1}\def\cite#1{{#1}}\def\[#1]{[\cite{#1}]}

\def\(#1){(\call{#1})}
\def\call#1{{#1}}\def\taghead#1{{#1}}

\def\references{\head{REFERENCES}\beginparmode\frenchspacing\parskip=0pt}
\gdef\refis#1{\item{#1.\ }}
\gdef\journal#1,#2,#3,#4.{#1~{\bf #2}, #3 (#4)}
\def\endreferences{\body}
\def\endit{\endmode\vfill\supereject}\let\endpaper=\endit

\def\gsim{\mathrel{\raise.3ex\hbox{$>$\kern-.75em\lower1ex\hbox{$\sim$}}}}
\def\lsim{\mathrel{\raise.3ex\hbox{$<$\kern-.75em\lower1ex\hbox{$\sim$}}}}
\def\square{\kern1pt\vbox{\hrule height 0.6pt\hbox{\vrule width 0.6pt\hskip 3pt
   \vbox{\vskip 6pt}\hskip 3pt\vrule width 0.6pt}\hrule height 0.6pt}\kern1pt}
\def\sla{\raise.15ex\hbox{$/$}\kern-.72em}

\def\iafedir{Instituto de Astronom\'\i a y F\'\i sica del Espacio\\Casilla de
          Correo 67 - Sucursal 28, 1428 Buenos Aires -- Argentina}
\def\fceyn{Departamento de F{\'\i}sica\\
      Facultad de Ciencias Exactas y Naturales, Universidad de Buenos Aires\\
      Ciudad Universitaria - Pabell\'on I, 1428 Buenos Aires, Argentina}

\catcode`@=11
\newcount\r@fcount \r@fcount=0\newcount\r@fcurr
\immediate\newwrite\reffile\newif\ifr@ffile\r@ffilefalse
\def\w@rnwrite#1{\ifr@ffile\immediate\write\reffile{#1}\fi\message{#1}}
\def\writer@f#1>>{}
\def\referencefile{\r@ffiletrue\immediate\openout\reffile=\jobname.ref%
  \def\writer@f##1>>{\ifr@ffile\immediate\write\reffile%
    {\noexpand\refis{##1} = \csname r@fnum##1\endcsname = %
     \expandafter\expandafter\expandafter\strip@t\expandafter%
     \meaning\csname r@ftext\csname r@fnum##1\endcsname\endcsname}\fi}%
  \def\strip@t##1>>{}}

\def\citeall#1{\xdef#1##1{#1{\noexpand\cite{##1}}}}
\def\cite#1{\each@rg\citer@nge{#1}}
\def\each@rg#1#2{{\let\thecsname=#1\expandafter\first@rg#2,\end,}}
\def\first@rg#1,{\thecsname{#1}\apply@rg}
\def\apply@rg#1,{\ifx\end#1\let\next=\relax%
\else,\thecsname{#1}\let\next=\apply@rg\fi\next}%
\def\citer@nge#1{\citedor@nge#1-\end-}
\def\citer@ngeat#1\end-{#1}
\def\citedor@nge#1-#2-{\ifx\end#2\r@featspace#1
  \else\citel@@p{#1}{#2}\citer@ngeat\fi}
\def\citel@@p#1#2{\ifnum#1>#2{\errmessage{Reference range #1-#2\space is bad.}
    \errhelp{If you cite a series of references by the notation M-N, then M and
    N must be integers, and N must be greater than or equal to M.}}\else%
{\count0=#1\count1=#2\advance\count1 by1\relax\expandafter\r@fcite\the\count0,%
  \loop\advance\count0 by1\relax
    \ifnum\count0<\count1,\expandafter\r@fcite\the\count0,%
  \repeat}\fi}
\def\r@featspace#1#2 {\r@fcite#1#2,}    \def\r@fcite#1,{\ifuncit@d{#1}
    \expandafter\gdef\csname r@ftext\number\r@fcount\endcsname%
    {\message{Reference #1 to be supplied.}\writer@f#1>>#1 to be supplied.\par
     }\fi\csname r@fnum#1\endcsname}
\def\ifuncit@d#1{\expandafter\ifx\csname r@fnum#1\endcsname\relax%
\global\advance\r@fcount by1%
\expandafter\xdef\csname r@fnum#1\endcsname{\number\r@fcount}}
\let\r@fis=\refis   \def\refis#1#2#3\par{\ifuncit@d{#1}%
    \w@rnwrite{Reference #1=\number\r@fcount\space is not cited up to now.}\fi%
  \expandafter\gdef\csname r@ftext\csname r@fnum#1\endcsname\endcsname%
  {\writer@f#1>>#2#3\par}}
\def\r@ferr{\endreferences\errmessage{I was expecting to see
\noexpand\endreferences before now;  I have inserted it here.}}
\let\r@ferences=\references
\def\references{\r@ferences\def\endmode{\r@ferr\par\endgroup}}
\let\endr@ferences=\endreferences
\def\endreferences{\r@fcurr=0{\loop\ifnum\r@fcurr<\r@fcount
    \advance\r@fcurr by 1\relax\expandafter\r@fis\expandafter{\number\r@fcurr}%
    \csname r@ftext\number\r@fcurr\endcsname%
  \repeat}\gdef\r@ferr{}\endr@ferences}
\let\r@fend=\endpaper\gdef\endpaper{\ifr@ffile
\immediate\write16{Cross References written on []\jobname.REF.}\fi\r@fend}
\catcode`@=12
\citeall\refto\citeall\ref\citeall\Ref
\catcode`@=11
\newcount\tagnumber\tagnumber=0
\immediate\newwrite\eqnfile\newif\if@qnfile\@qnfilefalse
\def\write@qn#1{}\def\writenew@qn#1{}
\def\w@rnwrite#1{\write@qn{#1}\message{#1}}
\def\@rrwrite#1{\write@qn{#1}\errmessage{#1}}
\def\taghead#1{\gdef\t@ghead{#1}\global\tagnumber=0}
\def\t@ghead{}\expandafter\def\csname @qnnum-3\endcsname
  {{\t@ghead\advance\tagnumber by -3\relax\number\tagnumber}}
\expandafter\def\csname @qnnum-2\endcsname
  {{\t@ghead\advance\tagnumber by -2\relax\number\tagnumber}}
\expandafter\def\csname @qnnum-1\endcsname
  {{\t@ghead\advance\tagnumber by -1\relax\number\tagnumber}}
\expandafter\def\csname @qnnum0\endcsname
  {\t@ghead\number\tagnumber}
\expandafter\def\csname @qnnum+1\endcsname
  {{\t@ghead\advance\tagnumber by 1\relax\number\tagnumber}}
\expandafter\def\csname @qnnum+2\endcsname
  {{\t@ghead\advance\tagnumber by 2\relax\number\tagnumber}}
\expandafter\def\csname @qnnum+3\endcsname
  {{\t@ghead\advance\tagnumber by 3\relax\number\tagnumber}}
\def\equationfile{\@qnfiletrue\immediate\openout\eqnfile=\jobname.eqn%
  \def\write@qn##1{\if@qnfile\immediate\write\eqnfile{##1}\fi}
  \def\writenew@qn##1{\if@qnfile\immediate\write\eqnfile
    {\noexpand\tag{##1} = (\t@ghead\number\tagnumber)}\fi}}
\def\callall#1{\xdef#1##1{#1{\noexpand\call{##1}}}}
\def\call#1{\each@rg\callr@nge{#1}}
\def\each@rg#1#2{{\let\thecsname=#1\expandafter\first@rg#2,\end,}}
\def\first@rg#1,{\thecsname{#1}\apply@rg}
\def\apply@rg#1,{\ifx\end#1\let\next=\relax%
\else,\thecsname{#1}\let\next=\apply@rg\fi\next}
\def\callr@nge#1{\calldor@nge#1-\end-}\def\callr@ngeat#1\end-{#1}
\def\calldor@nge#1-#2-{\ifx\end#2\@qneatspace#1 %
  \else\calll@@p{#1}{#2}\callr@ngeat\fi}
\def\calll@@p#1#2{\ifnum#1>#2{\@rrwrite{Equation range #1-#2\space is bad.}
\errhelp{If you call a series of equations by the notation M-N, then M and
N must be integers, and N must be greater than or equal to M.}}\else%
{\count0=#1\count1=#2\advance\count1 by1\relax\expandafter\@qncall\the\count0,%
  \loop\advance\count0 by1\relax%
    \ifnum\count0<\count1,\expandafter\@qncall\the\count0,  \repeat}\fi}
\def\@qneatspace#1#2 {\@qncall#1#2,}
\def\@qncall#1,{\ifunc@lled{#1}{\def\next{#1}\ifx\next\empty\else
  \w@rnwrite{Equation number \noexpand\(>>#1<<) has not been defined yet.}
  >>#1<<\fi}\else\csname @qnnum#1\endcsname\fi}
\let\eqnono=\eqno\def\eqno(#1){\tag#1}\def\tag#1$${\eqnono(\displayt@g#1 )$$}
\def\aligntag#1\endaligntag  $${\gdef\tag##1\\{&(##1 )\cr}\eqalignno{#1\\}$$
  \gdef\tag##1$${\eqnono(\displayt@g##1 )$$}}
\def\eqalignno#1{\displ@y \tabskip\centering
  \halign to\displaywidth{\hfil$\displaystyle{##}$\tabskip\z@skip
    &$\displaystyle{{}##}$\hfil\tabskip\centering
    &\llap{$\displayt@gpar##$}\tabskip\z@skip\crcr
    #1\crcr}}
\def\displayt@gpar(#1){(\displayt@g#1 )}
\def\displayt@g#1 {\rm\ifunc@lled{#1}\global\advance\tagnumber by1
        {\def\next{#1}\ifx\next\empty\else\expandafter
        \xdef\csname @qnnum#1\endcsname{\t@ghead\number\tagnumber}\fi}%
  \writenew@qn{#1}\t@ghead\number\tagnumber\else
        {\edef\next{\t@ghead\number\tagnumber}%
        \expandafter\ifx\csname @qnnum#1\endcsname\next\else
        \w@rnwrite{Equation \noexpand\tag{#1} is a duplicate number.}\fi}%
  \csname @qnnum#1\endcsname\fi}
\def\ifunc@lled#1{\expandafter\ifx\csname @qnnum#1\endcsname\relax}
\let\@qnend=\end\gdef\end{\if@qnfile
\immediate\write16{Equation numbers written on []\jobname.EQN.}\fi\@qnend}
\catcode`@=12

\magnification 1200
\oneandahalfspace
\title{Chaotic Friedmann - Robertson - Walker Cosmology}
\author{ Esteban Calzetta and Claudio El Hasi}
\affil
\iafedir
and
\fceyn
\abstract
We show that the dynamics of a spatially closed
Friedmann - Robertson - Walker Universe conformally coupled to a real, free,
massive scalar field, is chaotic,
for large enough field amplitudes. We do
so by proving that this system is integrable under the
adiabatic approximation, but that the corresponding
KAM tori break up when non
adiabatic terms are considered. This finding is confirmed by
numerical evaluation of the Lyapunov exponents associated with
the system, among other criteria. Chaos sets strong limitations
to our ability to predict the value of the field at the Big Crunch,
from its given value at the Big Bang.
\bigskip
PACS 04.20.Cv, 98.80.Bp, 03.20.+i
\endtitlepage

{\bf I - Introduction}
\bigskip
In this paper, we shall present analytical and numerical evidence
of chaotic behavior in a cosmological model, consisting of a
spatially closed, Friedmann - Robertson - Walker (FRW) Universe, filled
with a conformally coupled but massive real scalar field. although
this model is far too simplified to be considered realistic, its
simplicity itself makes it an interesting testing ground for the
implications of chaos for cosmology, either classical, semi - classical
or quantum. Moreover, the occurrence of chaotic motion in this
example supports the conjecture of a farther reaching connection
between chaos at the classical level, particle creation at the semiclassical
one, and decoherence in the full quantum cosmological treatment
( see below ).

The research on chaotic cosmological models began with the work by
the Russian school\refto{Belinskii}
 and by C. Misner and coworkers \refto{Misner} on chaos in
Bianchi type IX vacuum cosmologies. although the hopes
initially placed on these models ( such as a resolution of the
horizon problem from the mixing effect associated to  chaos ) could
not be sustained, Bianchi IX is still by far the best studied
chaotic cosmology\refto{mixchaos}
, both because of its intrinsic interest, and as a
test bench for more general questions, such as whether chaos is
an observer dependent phenomenon\refto{Burd}. In the broader context of
General Relativity, chaos has been considered in connection
to geodesic motion, both in cosmological\refto{Balasz} and Black Hole space
times\refto{Contopoulos}.

The relative paucity of examples of relativistic chaos\refto{Buchler}
 makes
it hard to disentangle the general features ( if any ) of these
phenomena from the ``miracles'' proper to each peculiar manifestation,
such as the occurrence of the Gauss' map hidden in Bianchi IX
dynamics, which led to the discovery of cosmological chaos in the first
place\refto{Belinskii}
. It is therefore of the utmost importance to develop a systematic
search for instances of relativistic chaos, to enhance our battery
of examples, and therefore better to aim future research.

As pointed out in an earlier communication\refto{BHchaos}, the class of near
integrable systems (NIS) is an interesting field for such a search.
A NIS is an integrable system which becomes non integrable under
the effect of a perturbation. The non integrable perturbed dynamics
may or may not be, in turn, chaotic. A typical situation occurs
when the unperturbed dynamics displays an unstable fixed point,
asymptotically joined to itself by a nontrivial orbit ( the
so called ``homoclinic loop'' ). If the loop is destroyed by
the perturbation, then a ``stochastic layer'' forms in its neighborhood.
There are subsets in this layer where the dynamics is equivalent to
a Bernoulli shift\refto{Guckenheimer}
. This is the same degree of chaotic behavior
characteristic of the Bianchi IX example. A detailed discusion of
this scenario, geared to its application in relativistic problems,
has been given in Ref.\cite{BHchaos}.

The case at hand in this paper belongs to a wider class where there
are no homoclinic points in the unperturbed dynamics. However, in a
certain sense, the perturbation both creates them and destroys the
corresponding homoclinic loops, allowing the formation of stochastic
layers. This road to chaos results from the overlapping of several
resonances between the external perturbation and the unperturbed
motion. In the neighborhood of a resonant region in phase space,
one of these resonances dominates over the rest; because of this
dominant resonance, KAM tori are destroyed in this region. In
particular, the torus for which the resonance is exact  dissapears,
leaving behind a discrete set of both stable and unstable fixed
points\refto{ArnoldAv}.
The unstable fixed points are connected to each other through
doubly asymptotic orbits or separatrices; chaos occurs whenever
the separatrices are further destroyed by the effect of the
secondary resonant terms. In the neighborhood of a destroyed
separatrix, it is possible to find ( measure zero ) invariant sets
where the restricted dynamics is equivalent to a Bernoulli flow.
More generally, chaotic behavior, under the form of extreme
sensitivity regarding initial conditions, is displayed in a set
of full measure around the separatrix, the so called stochastic layer
\refto{Chirikov}.

The particular model we shall present here has been chosen both
because of its simplicity and because of its relevance to
the discussion of the ties between classical and quantum
cosmological models\refto{Berry}. Indeed, similar models have been used by
Hawking and Page\refto{HPage} to discuss the relationship between the
cosmological and thermodynamic arrow of time, in the framework
of Quantum Cosmology. Since in this model the system behaves
as a Bernoulli flow in appropiate invariant sets, time
reversal invariance
may be spontaneously broken, and thus a ``chaotic'' arrow of
time appears\refto{Elskens}. The behavior of this arrow supports Page's picture
of the relationship between the cosmological and the thermodynamic
ones.

In the cosmological literature more generally, fundamental scalar
fields are usually considered within the framework of Inflationary
models. In these models, however, minimal coupling is often preferred
to conformal one\refto{Abbott}. Thorough analysis of the dynamics of a
minimally
coupled field in a FRW background have been developped by several
authors\refto{mincoup}; they have found that Inflationary cosmologies act as an
attractor in phase space, but little or no evidence of chaos.
Further studies of non minimally coupled fields in open universes
have yielded similar results\refto{nonmin}. In
the conformally coupled model we shall present here, on the other hand,
``inflationary'' periods, where the radius $a$ of the Universe increases
by several orders of magnitude, are a common feature of the
solutions to the full dynamics. In this model, ``inflation'' is not powered
by an effective cosmological constant, but rather by the effective
Newton's constant becoming negative for large values of the field
\refto{Kiefer}.

Another source of interest in this model is that it is one of the simplest
cosmologies where particle creation by the gravitational field occurs
\refto{Parker}.
Indeed, particle creation in models of this kind has been analyzed
by Hartle and others
\refto{Hartle}. In a perturbative analysis, the spectrum of
created particles is closely related to the Fourier decomposition
of $a^2(\eta )$, where $\eta$ stands for conformal time,
and $a$ is the ``Radius of the Universe'' ( see
 below ). This fact, and the connection between particle creation
and the breakdown of the WKB approximation through Stokes' phenomenon
\refto{me},
suggest that particle creation is indeed the effect of resonances
between the evolution of $a$ and that of the scalar field. If this
were true, then a strong correlation between classical chaos and
semiclassical particle creation should exist. The relationship
of semiclassical particle creation to quantum cosmological
decoherence has been discussed elsewhere\refto{Diego}.

In this paper we shall investigate our cosmological model both through
analytical (perturbative) and numerical (non perturbative)
methods. After showing how the full Hamiltonian can be split
into an integrable part plus a perturbation, we shall show
how the KAM tori of the integrable part are destroyed and
replaced by stochastic layers under the effect of the perturbation.
We shall analyze the onset of chaos through three methods of
increasing sophistication. We shall begin by employing
Chirikov's resonance overlap criterium\refto{Chirikov} to show that the
stochastic layer of a given resonance is wide enough to intrude
into the layers of its neighbors. Then we shall employ the
Melnikov Integral\refto{Guckenheimer}
 to show that the separatrix created by each
resonance is split by the effect of the secondary perturbations.
Finally, we shall investigate the dynamics close to a separatrix
by constructing the corresponding separatrix or standard map
\refto{Chirikov}.
We shall use this map to discuss the important issue of whether
the sensitive dependence on initial conditions proper to chaos
is strong enough to lead to observable effects within the lifespan
of the Universe. It should be remembered that Bianchi IX models
have been found lacking in this regard\refto{Misner}.

Numerical analysis will allow us to go beyond perturbation
theory. We shall present results from numerical integration of the
full equations of motion, both in the chaotic and non chaotic
regimes; plots of the field against $a$, where the change in the
topology of the orbits, subsequent to chaos, can be observed;
numerical estimates of the Lyapunov coefficients\refto{Wolf}
in the stochastic layers; and finally, a plot of the values
of the field at the Big Crunch vs. those at the Big Bang, for
fixed initial geometry, to demonstrate the loss of predictibility
subsequent to chaos.

The paper is organized as follows. We introduce the model in the
following section, where we also develop its perturbative analysis.
Our numerical results are presented in Section III. Finally, we
briefly state our conclusions in Section IV.
\bigskip
{\bf II - The model and its perturbative treatment}
\bigskip
{\it II.1 - The model}
\medskip
Let us begin by introducing the model and how we shall split
its Hamiltonian into an unperturbed part plus a perturbation.

As already discussed, our cosmological model assumes a FRW
spatially closed geometry, that is, a line element

$$ds^2=a^2(\eta )[-d\eta^2+d\chi^2+\sin^2\chi (d\theta^2+\sin^2
\theta d\varphi^2)]\eqno(2.1)$$

Where $0\le\varphi\le 2\pi$, $0\le\theta\le\pi$, $0\le\chi\le\pi$,
and $\eta$ stands for ``conformal'' time. For concreteness, we
shall consider only models starting from a cosmic singularity,
that is, we restrict $\eta$ to be positive, with $a(0)=0$.
Also, as indicated by the dynamics, we shall assume that after
the Big Crunch ( that is, when $a$ returns to $0$ ), a new
cosmological cycle begins, now with $a\le 0$. Therefore, a
complete periodic orbit describes the birth and death of two
Universes.

 The
gravitational dynamics is described by the Einstein - Hilbert
action

$$S_g=\int~d^4x~\sqrt{-g}m_p^2R\eqno(2.2)$$

Where the determinant of the metric $-g=a^4\sin^2\chi
\sin\theta$, and the scalar curvature

$$R=6[{\ddot a\over a^3}+{1\over a^2}]\eqno(2.3)$$

(We shall use MTW conventions throughout
\refto{MTW}) A dot represents a $\eta$
derivative, and
$m_p$ is Planck's mass. For
simplicity, we shall assume $m_p=\sqrt{1/12v}$, where $v=2\pi^2$ is
the conformal volume of an spatial surface.

The action for a conformally coupled, massive, real scalar field
is given by

$$ S_f=\int~d^4x~\sqrt{-g}~({-1\over 2})[\partial_{\mu}\Phi\partial^{\mu}
\Phi +(m^2+(1/6)R)\Phi^2]\eqno(2.4)$$

Where $m^2$ is the mass. Consistency with the symmetries of the
background geometry demands the field be homogeneous. Parametrizing
the field as $\Phi =\phi/v^{(1/2)}a$, performing the spatial
integrals, and discarding total derivatives, we are led to a
dynamical system with two degrees of freedom, $a$ and $\phi$,
and Hamiltonian

$$H=({1\over 2})[-(\pi^2+a^2)+(p^2+\phi^2)+m^2a^2\phi^2]\eqno(2.5)$$

Where $\pi$ and $p$ are the momenta conjugated to $a$ and $\phi$
respectively. At this point we must recall that, because we have
relinquished our gauge freedom in writing the line element as in Eq.\(2.1),
we are missing one of Einstein's equations, namely, the Hamiltonian
constraint\refto{MTW}. We reintroduce this constraint as a restriction on
allowable initial conditions

$$H=0\eqno(2.6)$$

Which is clearly respected by the dynamics.

When $m^2=0$, the Hamiltonian \(2.5) is obviously integrable; for
nonvanishing $m^2$, this is no longer so obvious, and we must
resort to perturbative methods ( or solve the dynamics numerically,
see next section ). It is tempting to consider the massless
Hamiltonian as the unperturbed one, with the last term in Eq. \(2.5)
as perturbation. This is questionable, however, on the grounds that
for ``macroscopic'' Universes, very easily we obtain $m^2a^2\gg 1$.
For example, for our own Universe $a\sim 10^{60}$, and for a mass
of 1 eV, we get $m\sim 10^{-28}$, so $ma\sim 10^{32}$. In this regime, the
last term in \(2.5) is in no way small compared with the other ones.

We shall therefore proceed in a different way. Let us observe that,
if the evolution of $a$ is slow compared to the oscillations in $\phi$,
then its main effect is to produce an adiabatic change in the frequency
of the latter. We therefore introduce the ``adiabatic'' amplitude and phase
$j$ and $\varphi$

$$\phi =\sqrt{{2j\over\omega}}\sin\varphi\eqno(adamp)$$

$$p=\sqrt{2\omega j}\cos\varphi\eqno(adang)$$

Where $\omega^2=1+m^2a^2$ is the instantaneous frequency of the field.
This transformation is canonical; it can be accomplished by means
of the generating functional

$$S_1=Pa+({\omega\phi^2\over 2\tan\varphi})\eqno(S1)$$

But, because of the $a$ dependence in $\omega$, we are forced to
change the geometrical momentum from $\pi$ to $P$, according to

$$\pi =P+{m^2aj\over 2(1+m^2a^2)}\sin 2\varphi\eqno(pitoP)$$

We now substitute the new variables in the Hamiltonian, and
rewrite this as

$$H=-(H_0+\delta H)\eqno(sign)$$

where the unperturbed
Hamiltonian

$$H_0=({1\over 2})[P^2+a^2]-j\sqrt{1+m^2a^2}\eqno(unph)$$

is obviously integrable ( $H_0$ and $j$ are constants of motion in
involution ), and the perturbation

$$\delta H={m^2aPj\over 2(1+m^2a^2)}\sin 2\varphi
+[{m^2aj\over 4(1+m^2a^2)}]^2(1-\cos 4\varphi )\eqno(perth)$$

remains small both for small and large Universes.
\bigskip
{\it II.2 - Solving the unperturbed dynamics}
\medskip

The dynamics of $a$, as generated by $H_0$, is obviously bounded. The
point $a=0$ is a fixed point; it is stable if $m^2j\le 1$, , and
unstable otherwise. In this second case, there is an homoclinic loop
associated with it. However, this orbit does not satisfy the
Hamiltonian constraint, Eq. \(2.6); rather, we have $H_0=-j$ on the
homoclinic loop.

The equations of motion are simpler if written in terms of
a new variable $X=\sqrt{1+m^2a^2}$, rather than $a$ itself. The
transformation is canonical if we associate to $X$ the momentum

$$P_X={XP\over m\sqrt{X^2-1}}\eqno(PX)$$

In terms of the new variables $(X,P_X)$, the unperturbed Hamiltonian
reads

$$H_0=[{m^2(X^2-1)\over 2X^2}]P_X^2+({1\over 2m^2})[(X-m^2j)^2-(m^2j)^2-1]
\eqno(hamnev)$$

For fixed $H_0=h$ and $j$, provided $h\ge -j$,
 $X$ ranges from $1$ to the classical turning point
$X_T=m^2j+m\sqrt{K}$, where $K={(1/m^2)+(mj)^2+2h}$ ( The reasons for the
somewhat unusual notation shall be clear below).

Although the Hamiltonian Eq. \(hamnev) can be explicitly integrated
( the solution involves the use of elliptic integrals
\refto{Whittaker}), for  our
purposes it will be better to concentrate on a particular case, where
a number  of  simplifications  will  be  available.    Concretely, we shall
consider the large $j$ limit, with $H_0\sim 0$ and $m$ fixed. In this
limit, we find $X\gg 1$ for most of the orbit. Indeed, from Hamilton's
equations, we get $\dot X\sim 2m\sqrt{h+j}\sqrt{X-1}$ when $X\sim 1$
and $H_0=h$. Therefore, even if $X$ starts at $1$, $X-1$ becomes of order
unity after a lapse $\delta t\sim (m\sqrt{h+j})^{-1}$, which is small in
the case under consideration. Now, for $X\gg 1$, the Hamiltonian
Eq. \(hamnev) simplifies to

$$H_0=({m^2\over 2})P_X^2+({1\over 2m^2})[(X-m^2j)^2-(m^2j)^2-1]
\eqno(rehamnev)$$

If we parametrize

$$P_X={\sqrt{K}\over m}\cos 2\alpha\eqno(traction)$$

$$X=m^2j+m\sqrt{K}\sin 2\alpha\eqno(braction)$$

Eq. \(rehamnev) reduces to

$$H_0=({1\over 2})[K-(mj)^2-{1\over m^2}]\eqno(contrahamnev)$$

As $X$ goes from $1$ to $X_T$, that is, over a quarter
of an orbit, $\sin 2\alpha$ goes from $-(m^2j-1)/m\sqrt{K}$ to $1$.
For large $j$ and $H_0\sim 0$, however, $K$ is very close to $(mj)^2$.
So the end points can be taken as $\alpha\sim\pm\pi /4$, in which case
$K$ is precisely the associated action variable. We have succeeded
in integrating the unperturbed motion, only that, because the
reparametrization eqs. \(traction) and \(braction) involves $j$,
the angle canonically conjugated to it is no longer $\varphi$,
but

$$\theta =\varphi -m\sqrt{K}\cos 2\alpha\eqno(theta)$$

The transformation eqs. \(traction), \(braction) and \(theta), from
variables $(X, P_X, \varphi, j)$ to new variables $(\alpha , K, \theta,
j'=j)$ is canonical; indeed, it is generated by

$$S_2=j'\varphi +{(X-m^2j')^2\over 2m^2\tan 2\alpha}\eqno(genfunc)$$

In the following, we shall omit the prime on $j'$.
\bigskip
{\it II.3 - Analysis of the perturbation}
\medskip
Having reduced the unperturbed Hamiltonian Eq. \(unph) to action - angle
form \(contrahamnev), let us return to the perturbation, Eq. \(perth).
We notice, first of all, that the first term in the perturbation
dominates the second ( except close to the turning points ). For our
purposes, it shall be sufficient to retain only the dominant
perturbation.

We also notice that, $m\sqrt{K}$ being large,
$\sin 2\varphi\equiv
\sin 2(\theta +m\sqrt{K}\cos 2\alpha)$
(cfr. Eq. \(theta)) is a strongly oscillating function
of $\alpha$. Thus, in computing the Fourier expansion of the
perturbation, we may substitute the non oscillatory factor
$m^2aPj/2(1+m^2a^2)$ by its mean value over a quarter orbit, which is
easily found to be $(j/\pi )[\ln X_T-(X_T^2-1)/2X_T^2]\sim (j\ln X_T)/\pi$.

The Fourier expansion of the oscillatory factor itself is

$$\sin 2(\theta +m\sqrt{K}\cos 2\alpha)=J_0(2m\sqrt{K})\sin 2\theta
+\sum_{n=1}^{\infty}
J_n(2m\sqrt{K})[\sin\delta_n^+
+\sin\delta_n^- ]\eqno(Fourier)$$

Where $J_n$ is the usual Bessel function, and
$\delta_n^{\pm}=(2\theta \pm 2n(\alpha+(\pi /4)))$. Therefore, resonances occur
whenever

$$\omega_j\pm n\omega_K=0\eqno(rescon)$$

$\omega_{j,K}$ being the frequencies associated to $\theta$ and $\alpha$,
respectively. From Eq.\(contrahamnev), we have $\omega_j\sim m^2j$
( we have reinstated the proper sign, cfr Eq.\(sign) ), and $\omega_K
\sim -1/2$. We thus find a tower of resonances, corresponding to
all positive values of $n$;  the n-th resonance occurs at $j_n\sim n/2m^2$,
independently of $K$. If we further impose the Hamiltonian constraint,
then $K$ must take the value $K_n\sim (mj_n)^2=(n/2m)^2$.

The n-th resonant term in the Hamiltonian, with its proper sign, reads

$$({-j\over \pi})(\ln X_T)J_n( 2m\sqrt{K})\sin (2\theta +2n(\alpha+(\pi /4)))
\eqno(nres)$$

To analyze the perturbed motion it is sufficient to approximate
the prefactor by its value at resonance. So doing,
and using the proper asymptotic form for the Bessel
function\refto{Courant}, Eq.\(nres)
reduces to

$$({-\epsilon\over m^2})n^{2/3}(\ln n)\sin (2\theta +2n(\alpha+(\pi /4)))
\eqno(apnres)$$

where $\epsilon$ is a numerical coefficient, $\epsilon\sim 0.111827...$
\vfill
\eject
{\it II.4 - Solving the perturbed motion}
\medskip
To analyze the motion in the presence of the perturbation, let
us focus first on the neighborhood of the $n$-th resonance ( the
calculations below are simplest if $n=4k+1$ for some $k$, which we shall
assume ). Let us keep only the dominant resonant term Eq. \(apnres)
in the Hamiltonian, and introduce the new action variables
$\xi =(j-j_n)/2$ and $\kappa =-(K-K_n-n(j-j_n))/2$, which
vanish at the resonant point. These variables
are canonically conjugated to the angles
$\psi =2(\theta +n\alpha )$ and $\tau=-2\alpha$, respectively
( the transformation is generated by
$S_3=[K_n-2(\kappa-n\xi )]\alpha +[j_n+2\xi ]\theta$ ). In terms
of the new variables, the resonant Hamiltonian becomes
( cfr. Eqs. \(contrahamnev) and \(apnres) )

$$H_n=\kappa +2m^2\xi^2-({\omega_0^2\over 4m^2})\cos\psi\eqno(nresham)$$

Where $\omega_0^2=4\epsilon n^{2/3}\ln n$.

Because the Hamiltonian \(nresham) is linear in $\kappa$, it can be
considered as resulting from the ``parametrization'' of a one
degree of  freedom  system\refto{Kuchar}.
To  analyze  the  resulting  motion,  it  is
convenient to ``deparametrize'' it, that is, to promote $\tau$ to
the roll of ``time''. The evolution of the ``true'' degree of freedom
$\psi$ as $\tau$ unfolds is described by the Hamiltonian $E=-\kappa$.

{}From Eq. \(nresham) and the Hamiltonian constraint \(2.6), it is obvious
that $E$ is simply the Hamiltonian of a non linear pendulum, $\omega_0$
being the frequency of small oscillations around the stable equilibrium
point $\psi =0$. The pendulum also has an unstable equilibrium point
$\psi =\pm\pi$, joined to itself by a separatrix.

Following Chirikov\refto{Chirikov}, we define the ``width''
of the resonance as
the maximum value of $\xi$ along the separatrix, $\Delta\xi\sim
\omega_0/2m^2$. Since $\omega_0\sim n^{1/3}$, it is clear that
for large $n$ each resonance is much wider that the separation
$1/2m^2$ between resonances. Thus, according to Chirikov's
criterium, the behavior of the perturbed system must be
chaotic.

To obtain a more detailed picture, let us add to the resonant
Hamiltonian Eq. \(nresham) the first secondary resonance ($n=4k+2$),

$$\delta H_n\sim (\omega^2_0/4m^2)\sin (\psi -\tau )\eqno(deltahn)$$

( The same
term is added to the ``deparametrized'' Hamiltonian $E$). We can
now apply the criterium that chaos will occur if the secondary
resonance is able to destroy the separatrix created by the
primary resonance. This can be determined by computing the total change
in the unperturbed pendulum Hamiltonian, induced by the perturbation,
as the pendulum swings along the separatrix, which is given
by the Melnikov Integral\refto{Guckenheimer}

$$I=-\omega_0^2\int_{-\infty}^{+\infty}~d\tau~\xi\cos (\psi -\tau )\eqno
(Melnikov)$$

In our case\refto{Chirikov},
$I=-\Delta E\cos\psi_0$, where $\psi_0$ is the value of $\psi$
at $\tau =0$, and

$$\Delta E={\pi e^{{\pi\over 2\omega_0}}\over m^2\sinh{\pi\over\omega_0}}
\eqno(deltaE)$$

The fact that $I$ displays isolated zeroes as a function of $\psi_0$
is again proof of the presence of chaos when the perturbation is turned on
\refto{Guckenheimer}.

\bigskip
{\it II.5 - The Separatrix Map}
\medskip
The oscillatory behavior of the Melnikov Integral proves implicitly
that the dynamics of our system, restricted to suitable invariant sets,
is equivalent to a Bernoulli flow. By constructing the so - called
``separatrix map''\refto{Chirikov},
we shall be able to show explicitly some of those
sets, and therefore acquire a much more immediate insight on the
evolution of the Universe near a resonance.

To build the separatrix map we shall adopt the picture, emerging from
the previous subsection, of the dynamics of the conformal field as
that of a pendulum, with angle $\psi$ and conjugated momentum $\xi$,
subject to the perturbation Eq.\(deltahn).

Let us suppose that, at a certain
instant $\tau_r$, the ``pendulum'' has energy $E_r$ and coordinate $\psi_r$.
If we follow the unperturbed motion over a closed orbit, the energy
will  keep   constant,  while  the  phase  $\varphi  =\tau  -\psi$  of  the
perturbation  shall  increase    to    $\varphi_{r+1}=\varphi_r+2\pi/\omega
(E_{r})$. Observe that the phase shift is independent of the phase.

If we consider the effect of the perturbation, over the same lapse,
we observe that $E$ does not remain constant. For motion close to the
separatrix, the change in $E$ is approximately the same than at the
separatrix itself, and so we find

$$E_{r+1}=E_r-\Delta E\cos\varphi_r\eqno(Esepmap)$$

The energy shift induces changes in the charactheristic
frequencies $\omega (E)$ of the unperturbed  orbits, and thus the
total phase shift is also affected. As a first approximation, we
may write

$$\varphi_{r+1}=\varphi_{r}+{2\pi\over\omega (E_{r+1})}\eqno(fsepmap)$$

Eqs. \(Esepmap) and \(fsepmap) define the separatrix map. Observe
that the phase shift is no longer independent of the phase.

The separatrix map preserves areas.  It also has a double sequence of fixed
points $x^{\pm}_k=(E_k,\pm\pi /2)$, where $k$ is an integer, and
$\omega (E_k)=1/k$. For large $k$, the fixed points approach the
separatrix, where\refto{Chirikov}
  $\omega  (E)\sim\pi\omega_0/\ln (32 E_0/\vert E_0-E\vert)$,
  $E_0\equiv\omega_0^2/4m^2$. Restricting ourselves
to motion below the separatrix, for simplicity, we find
$E_0-E_k\sim    32E_0~{\rm exp}(-k\pi\omega_0)$.

The behavior of the separatrix map close to a fixed point is determined
by its linear part.  It can be shown that all $x_k^+$ fixed points are
hyperbolic, while the $x_k^-$ points are hyperbolic only for large enough
$k$. For our purposes, it is sufficient to concentrate on the
properties of the map near the $x^+$ fixed points.

For large enough $k$, the eigenvalues of the linearized map around
$x_k^+$ are $\lambda_k$ and $1/\lambda_k$, where

$$\lambda_k\equiv\Delta               E~{d\over        dE}({2\pi\over\omega
(E)})\vert_{E_k}\eqno(avl)$$

For large $k$, $\lambda_k\sim (\Delta E/16\omega_0E_0){\rm exp}(k\pi\omega_0)$
diverges exponentially. The map is stretching in the direction
$v^+_k=\cos\beta_+\partial_{\varphi}+\sin\beta_+\partial_E$, where
$\tan \beta_+\sim\Delta E/(\lambda_k+1)$, and contracting in
the direction $v^-_k=-sin\beta_-\partial_{\varphi}+\cos\beta_-\partial_E$,
where $\tan\beta_-\sim (1/\Delta E)(1-(1/\lambda_k))$.

We can now see chaos arising. A small parallelogram, with a vertex
on $x_k^+$ and sides along the directions $v^+$ and $v^-$, is stretched
along the former, and contracted along the latter. Since $v^+$ is
essentially the $\varphi$ direction, the angular span of the original
parallelogram is dilated to many times $2\pi$. After modulo $2\pi$
identification, therefore, the image  is superimposed to the original
parallelogram, in a way essentially equivalent to Arnold's ``Cat
Map''\refto{ArnoldAv}.
By a standard procedure, then, it is possible to identify,
in our parallelogram, an invariant ( null ) set where the
separatrix map is Bernoulli\refto{Guckenheimer}.

Clearly, the procedure works only if the map is strongly dilating.
As a ballpark estimate, we can place the ``border of chaos'' at
the value of $E_k$ for which $\lambda_k\sim 1$. Thus, we are led to
estimate the width of the stochastic layer around the
departing separatrix at $\delta E\sim 2\Delta E/\omega_0$. Deep
in the resonance region, that is , for the $n$th resonance, $n$
large, Eq.\(deltaE) yields $\Delta E\sim\omega_0/m^2$ ( recall
that $\omega_0\sim n^{1/3}$ is large itself ), and so the
width of the stochastic layer approaches a constant value
$\delta E\sim 2/m^2$. This value is about four times the
distance between the resonances themselves, so the stochastic
layers merge and a stochastic sea is formed: any initial condition
in this region falls within the stochastic layer of some resonance
\refto{Chirikov}.

Almost more important to us, these results show that the loss of
information associated to chaotic behavior is observable within
the span of a single Universe. Indeed, we have identified chaos
in the oscillations of the scalar field, parametrized against
the angle variable associated to the radius of the Universe. Now,
over most of the stochastic layer, the field performs many oscillations
in the lapse between the Big Bang and the Big Crunch. This results
from the  fact  that  the  frequency  associated  with  the  expansion  and
contraction of the Universe is nearly constant for large $m^2j$,
while the frequencies associated to the field oscillations scale
as $\omega_0\sim n^{1/3}$. For example, at the outer
rim of the stochastic layer, the frequency of field oscillations
is $\sim (\pi/2)(\omega_0/\ln 2\omega_0)$, much larger than the
cosmological frequency. To reach field frequencies comparable to the
cosmic one, we must approach the separatrix, up to energies
$E_0-E\sim 32E_0~{\rm exp}(-2\pi\omega_0)$. Since the width
of the stochastic layer does not shrink with $n$, we see that the
fraction of it where the frequency of field oscillations
is comparable to the rate of
cosmic expansion and contraction, is negligible for deep resonances.

To summarize the content of this section, we have shown that for our
simple cosmological model, the Hamiltonian can be split into an
integrable and a non integrable part, the integrable one being the
main determinant of the dynamics both for small and large Universes.
Using perturbative methods, we have shown that the perturbation destroys
the invariant tori of the unperturbed motion. For $m^2j\gg 1$,
$j$ being the adiabatic amplitude of field oscillations, a stochastic
sea is formed, through the merger of the stochastic layers of the
individual resonances. Chaos manifest itself through seemingly random phase
shifts in the field oscillations as the Universe expands and
recollapses. Thus severe limitations to our ability to predict the
future behavior of the field should arise, even within the span of
a single Universe.
\bigskip
{\bf III - Numerical treatment of the model}
\bigskip
although the analysis in the previous section amounts to
rather impressive evidence of chaotic behavior in our model, it
suffers from the limitations of the perturbative approach
we have chosen. By solving the model numerically, we shall be
able to go beyond perturbation theory, and thus find an
independent confirmation of the results above.

In our numerical work, we have used the ``physical'' variables
$(a,\pi ,\phi ,p)$, whose equations of evolution follow from
the Hamiltonian Eq. \(2.5); the simplicity of these equations
makes this approach more appealing than other, more
sophisticated, alternatives.

To solve the model, we have used a Runge Kutta\refto{ReyPastor}
$5^{th}$ order routine, implemented on an IBM compatible PC 486.
As a check on the numerical code, in all runs we have surveyed
the value of the Hamiltonian, finding that it never exceeded
a threshold of $10^{-13}$. For simplicity, we shall set $m=1$
throughout.

Figures (1) and (2) show the evolution of a ``typical Universe''
in the region of ``weak chaos'' ($m^2j\sim .5$,
in the notation of Section II.1). Figure (1)
shows the evolution of the scale factor, and Figure (2) that of the
field. The nonlinearity is clearly visible, toghether with a
sharp rise of the amplitude associated to the radius of the Universe
in the last oscillations. Observe the simultaneous change in the
frequency of field oscillations.

In what follows, we shall subject our model to a series of numerical
experiments, searching for unambiguous signals of chaotic, rather
than complicated, behavior. To this end, we shall study its Lyapunov
exponents, the projection of an orbit on configuration space, and
finally the relationship between the values of the field at two
consecutive zero crossings of the radius of the Universe.
\bigskip
{\sl III.1 - Lyapunov exponents. May chaos be quantified?}
\medskip
It is possible to obtain a quantitative measure of chaos, through
the determination of Lyapunov exponents. These are defined by
considering the deformation of a $\epsilon$-ball of initial
conditions in phase space, in an ellipsoid of principal axis
$\lambda_i$ at time $t$. If we order the axis by size, then
the $i$ -th Lyapunov exponent is given by the double limit
$L_i={\rm lim}_{\epsilon\to 0}{\rm lim}_{t\to\infty}((\ln\lambda_i)/t)$.
In a Hamiltonian system, the Lyapunov exponents add to zero,
as a consequence of Liouville theorem; if the system, moreover,
is conservative, then two of the Lyapunov exponents are zero,
reflecting the invariance of the energy shells,
and the homogeneity of time. Thus, in a
conservative system with two degrees of freedom, such as ours,
only the largest Lyapunov exponent carries meaningful information.

While a rigorous determination of Lyapunov exponents is usually
numerically prohibitive, it is possible to estimate
them, following a proposal by Wolf {\sl et al.}\refto{Wolf}. The
idea is to construct  ``local'' Lyapunov exponents by taking
the average over time of the logarithms of the
eigenvalues of the linearized evolution operator. While these
``local'' Lyapunov exponents show a marked transient behavior
over short times, they approach the true Lyapunov exponent as
time goes to infinity.

Figures (3), (4) and (5) show plots of the largest Lyapunov
exponent for three different ``cosmologies'', corresponding
to $m^2j\sim .5$, $50$ and $5,000$, respectively. In the
latter cases, we have cut off the early transient, to
allow for a larger vertical scale. Beyond the transient regime, it
can be seen that the greatest Lyapunov exponents are clearly
positive, taking values in the intervals from $.4$ to $.8$,
$2$ to $3$, and $3.5$ to $4$, respectively. This is consistent
with the results of Section II.
\bigskip
{\sl III.2 - Orbits in configuration space. May chaos be proven?}
\medskip
In spite of their usefullness to the determination of chaos, Lyapunov
exponents are sometimes schewed in studies of relativistic
problems, since they are generally gauge dependent
\refto{Burd}. More reliable information can be obtained from
the analysis of Poincare sections. A Poincare section is the set
of points in which a given orbit of the system intersects,
in a given direction, a given plane in phase space. If the
motion is regular, the Poincare sections shall fit in smooth
curves, while this will not occur for chaotic motion.

Similarly, we can study the projection of a trajectory of
our system on the $(a,\phi )$ plane. If a second constant
of motion $C$, in involution with the Hamiltonian $H$,
existed, making the system integrable\refto{Arnold}, then at
every point of the trajectory it would be possible to
solve the equations $C={\rm constant}$ and $H={\rm constant}$
for $\pi$ and $p$, and thereby reduce Hamilton's equations
to parametric equations for a curve $\phi =\phi (a)$. For
example, if $m=0$, then we may take $C=(p^2+\phi^2)/2$. In
these circumstances, the projection of the orbit on the
$(a,\phi )$ plane fits into a smooth curve. Faillure
to do so, on the other hand, evidences  the non existence
of a second constant of motion ( disregarding pathological
cases ).

Figures (6), (7) and (8) display plots of $\phi$ vs $a$, again for
$m^2j$ equal to .5, 50 and 5,000, respectively. It can be seen
that, while in the first case it seems possible to fit the resulting
graph into a smooth Lissajous curve, this is no longer so in the
latter cases. This again confirms the results from Section II.
\bigskip
{\sl III.3 - Field values at consecutive cosmic singularities.
May chaos be observed?}
\medskip
In the above numerical tests, we have allowed the radius of the
Universe to cross zero several times. However, in the ``real''
world our observations cannot be projected beyond the cosmological
singularity. Thus it is crucial to determine whether the loss
of accuracy in the prediction of the future behavior of orbits,
associated to chaos, is strong enough to lead to observable
results within a single Universe.

In order to do that, we analyzed what happens with Universes
starting from the same geometrical condition ( that is,
same $a$ and $\pi$ ) but different values of $\phi$ ( and
$p$ given by the Hamiltonian constraint ). We begin our
computation a little while after the Big Bang, ending the
numerical calculation when $a$ becomes negative for the first time
( that is, on the Big Crunch ).  Repeating this procedure several
times, a plot of the final value $\phi_f$ of the field against the initial
one $\phi_i$ can be obtained. For regular behavior, this plot should allow
the prediction of $\phi_f$ given $\phi_i$ with the proper accuracy.

Figures (9), (10) and (11) show the results of numerical experiments
such as described, once more for $m^2j\sim .5,~50~ {\rm and}~ 5,000$,
respectively. In the plot Fig. (9), all simulations started at $a=.001$
and $\pi  =-1$,  while  $\phi_i$ ranged from $.001$ to $1$;  to obtain Fig
(10), we set the initial values of $a$ and $\pi$ at $.001$ and $-10$,
respectively, allowing $\phi_i$ to range from $.01$ to $10$; finally,
Fig.    (11)    corresponds    to   the  values  $a_i=.1$,  $\pi_i=-100$,
$.1\le\phi_i\le 100$.

The outcome of the simulations indicates
that, while $\phi_f$ is easily predicted
from $\phi_i$ in the first case, prediction becomes impossible
in the latter ones. Indeed, the correlation coefficient between $\phi_f$ and
$\phi_i$, which gives $.933$ for the data in Fig. (9), falls to $.049$ and
$.103$ for Figs. (10) and (11), respectively.
Moreover, regular
``islands'' which can still be seen in Fig (10) ( e. g., for
$\phi_i$ between .3 and .45 ) have quite dissapeared from Fig (11).

In conclusion, numerical simulation of the model strongly confirms
the results of the analysis in Section II. Moreover, chaos places
severe limitations in our ability to predict the future behavior
of the Universe, even before the Cosmic Singularity is reached
for the first time.
\bigskip
{\bf IV - Conclusions}
\bigskip
The analytical and numerical analysis above shows that the
dynamics of a spatially closed Friedmann - Robertson - Walker
Universe coupled to a conformal but massive scalar field
displays chaotic behavior. Indeed, in appropiate invariant
sets, the dynamics of this cosmological model is equivalent
to a Bernoulli shift, and thus essentially indistinguishable
from a purely random process\refto{Shields}. The analysis
above tells us how to identify those sets.

The random element we have alluded to
appears in the solutions of the model
through the nonlinear oscillations of the field as
the Universe evolves from the initial cosmic singularity
towards recollapse. As a result, our ability to predict the
future behavior of the field is limited. This phenomenon
demonstrably occurs within a single cosmic episode, and thus it is,
in principle, observable.

The interest of this result probably lies less in its direct
impact on present theories of the Universe, than in the avenues
it opens for further research. As a description of the Universe,
the model we have discussed here is obviously oversimplified.
More seriously, the chaotic behavior we have analyzed requires
high values of the field energy density, a regime where quantum
effects should not be disregarded.

In our view, the interest of this model lies in its
sheer simplicity. In this sense, we believe this model to be
an almost ideal ground where to investigate the general issues
associated with chaos and cosmology. For one thing, this model
is so simple that its analysis at the quantum and semiclassical
levels is not essentially harder than at the classical one.
This affords an excellent opportunity to discuss the semiclassical
limit of Quantum Cosmology in a highly non trivial framework, an
issue which we will discuss in a separate communication.

On the other hand, the model we have presented is far from
displaying the full richness of behavior than chaos
may bring to cosmology. We continue our research on this
most varied and rewarding field.
\bigskip
{\bf Acknowledgements}
\bigskip
We wish to thank all people who helped us to get started
in this project, or otherwise shared their know how with us,
specially to S. Blanco, A. Costa, G. Domenech, A. Fendrik, L.
Romanelli and N. Um\'erez.

This work was partially supported by the Directorate General for Science
Research and Development of the Comission of the European Communities,
under Contract ${\rm N}^o$ C11-0540-C, and by CONICET, UBA and
Fundaci\'on Antorchas.

\vfill
\eject
\references
\leftskip=30pt\parindent=-10pt
\def\new{\hfill\break\indent}\def\dash{---\hskip-1pt---}

\refis{Belinskii} Belinskii V A, Khalatnikov Z M and Lifshitz E M 1970
  Oscillatory approach to a singular point in the relativistic cosmology
  {\sl Adv. Phys.} {\bf19} 525-73 \new
Landau L and Lifschitz E M 1975 {\sl Classical Theory of Fields} (London:
  Pergamon)\new
Khalatnikov I M, Lifschitz E M, Khanin K M,
  Shchur L N, and Sinai Ya G 1985 On the stochasticity in relativistic
  cosmology {\sl J. Stat. Phys.} {\bf38} 97-114.

\refis{Misner} Misner C W 1970 Classical and quantum dynamics of a closed
  universe {\sl Relativity} ed. M Carmeli, S I Fickler and L Witten
  (New York: Plenum) 55-79 \new
\dash\ 1969a Quantum cosmology I {\sl Phys. Rev.} {\bf186} 1319-27 \new
\dash\ 1969b Absolute zero of time {\sl Phys. Rev.} {\bf186} 1328-33\new
Chitre D M 1972 Ph. D. Thesis, University of Maryland.\new
Ryan M 1972 {\sl Hamiltonian Cosmology} (Berlin, Springer - Verlag)\new
Ryan M and Shepley L 1975 {\sl Relativistic Homogeneous Cosmology}
( Princeton, Princeton University Press).

\refis{mixchaos}
Barrow J D 1982 Chaotic behaviour in general relativity {\sl Phys. Rep.}
  {\bf85} 1-49 \new
\dash\ 1987 {\sl Physics of Phase Space} ed. Y S Kim and W W Zachary
  (Berlin: Springer-Verlag) 18 \new
Barrow J D and Sirousse-Zia H 1989 Mixmaster cosmological models in
  theories of gravity with a quadratic Lagrangian {\sl Phys. Rev.} D {\bf39}
  2187-91\new
Zardecki A 1983 Modeling in Chaotic Relativity {\sl Phys. Rev.} D {\bf 28}
  1235-42\new
Francisco G and Matsas G E A 1988 Qualitative and numerical
  study of Bianchi IX models {\sl Gen. Rel. Grav.} {\bf20} 1047-54 \new
Burd A, Buric N and Ellis G 1990 A numerical analysis of chaotic behaviour
  in Bianchi IX models {\sl Gen. Rel. Grav.} {\bf22} 349-63 \new
Berger B K 1989 Quantum chaos in the mixmaster universe
  {\sl Phys. Rev.} D {\bf39} 2426-9 \new
\dash\ 1990 Numerical study of initially expanding mixmaster universes
  {\sl Class. Quantum Grav.} {\bf7} 203-16 \new
\dash\ 1992 Remarks on the Dynamics of the Mixmaster Universe
  {\sl Proceedings of the GR 13 International Conference} ed. C. Kozameh
  {\sl et al.} (to appear )\new
Burd A, Buric N and Tavakol R K 1991 Chaos, entropy and cosmology {\sl Class.
  Quantum Grav.} {\bf8} 123-33 \new
Pullin J 1991 Time and Chaos in General Relativity
  {\sl Relativity and Gravitation: Classical and Quantum}, ed. J. C. D'Olivo
  {\sl et al.} (Singapore,
  World Scientific) 189-97\new
Hobill D, Bernstein D, Wedge M and Simkins D 1991 The mixmaster cosmology
  as a dynamical system {\sl Class. Quantum Grav.} {\bf 8} 1155-71\new
Hobill D 1991 Sources of chaos in mixmaster cosmologies {\sl Non Linear
  Problems in Relativity and Cosmology}, ed. J Buchler, S Detweiler and
  J Ipser, Ann. N.Y. Acad. Sci. {\bf 631} (New York: N.Y. Acad. Sci.) 15-30\new
Rugh S E 1991 ( preprint )
  Chaos in the Einstein Equations {\sl NBI - HE } 91 - 59.

\refis{Burd} Burd A and Tavakol R K 1992 Gauge Invariance and Chaos
  {\sl Proceedings of the GR 13 International Conference} ed. C. Kozameh
  {\sl et al.} (to appear )\new
Tavakol R K 1992 Is General Relativity Fragile?
  {\sl Proceedings of the GR 13 International Conference} ed. C. Kozameh
  {\sl et al.} (to appear ).

\refis{Balasz} Balasz N L and Voros A 1986 Chaos in the pseudo sphere
  {\sl Phys. Rep.} {\bf143} 109 \new
Lockhart C M, Misra B and Prigogine I 1982 Geodesic instability and internal
  time in relativistic cosmology {\sl Phys. Rev.} D {\bf25} 921-9 \new
Tomaschitz R 1991 Relativistic quantum chaos in Robertson-Walker
  cosmologies {\sl J. Math. Phys.} {\bf32} 2571-9 \new
Gurzadyan V and Kocharyan A 1991, in {\sl Quantum Gravity} ed. M A Markov
  {\sl et al.} (Singapore: World Scientific).

\refis{Contopoulos}
Contopoulos G 1990 Periodic orbits and chaos around two fixed black holes. I
  {\sl Proc. Roy. Soc. (London)} {\bf A431} 183\new
\dash\ 1991a Periodic orbits and chaos around two fixed black holes. II
  {\sl Proc. Roy. Soc. (London)} {\bf A435} 551-62 \new
\dash\ 1991b Chaos around two fixed black holes, {\sl Non Linear
  Problems in Relativity and Cosmology}, ed. J Buchler, S Detweiler and
  J Ipser, Ann. N.Y. Acad. Sci. {\bf 631} (New York: N.Y. Acad. Sci.) 143-55

\refis{Buchler}
Buchler J R {\sl et al.} 1985 {\sl Chaos in Astrophysics} (New York: Reidel)
  \new
Buchler J R, Eichhorn H eds. 1987 {\sl Chaotic Phenomena in Astrophysics},
  Ann. N.Y. Acad. Sci. {\bf 497} (New York: N.Y. Acad. Sci.)\new
Buchler J R, Ipser J R, and Williams C A eds. 1988 {\sl Integrability
  in Dynamical Systems} Ann. N.Y. Acad. Sci. {\bf 536} (New York: N.Y.
  Acad. Sci.) \new
Buchler J R, Detweiler S L and Ipser J R eds. 1991 {\sl Non Linear Problems
  in Relativity and Cosmology} Ann. N.Y. Acad. Sci. {\bf 631} (New York:
  N.Y. Acad. Sci.).

\refis{BHchaos} Bombelli L and Calzetta E 1992 Chaos around a Black Hole
{\sl Class. Quantum Grav.} (to appear).

\refis{Guckenheimer} Guckenheimer J and Holmes P 1983 {\sl Non-Linear
  Oscillations, Dynamical Systems, and Bifurcations of Vector Fields}
  (Berlin: Springer-Verlag)\new
Wiggins S 1988 {\sl Global Bifurcations and Chaos}
  (Heidelberg: Springer-Verlag).

\refis{ArnoldAv} Arnold V I and Avez A 1968 {\sl Ergodic Problems of Classical
  Mechanics} (New York: Benjamin)

\refis{Chirikov} Chirikov B V 1979 A universal instability of many
  dimensional oscillator systems {\sl Phys. Rep.} {\bf52} 263 \new
Reichl L E and Zheng W M 1987 Non Linear Resonance and Chaos in
  Conservative Systems, in {\sl Directions in Chaos} ed. Hao Bai-Lin
  (Singapore, World Scientific)\new
Zaslavsky G M, Sagdeev R Z, Usikov D A and Chernikov A A 1991 {\sl Weak Chaos
  and Quasi Regular Patterns} (Cambridge: Cambridge University Press).

\refis{Berry} Berry M 1983 Semiclassical mechanics of regular and irregular
  motion {\sl Chaotic Behavior of Deterministic Systems}
  ed. G Iooss, R H G Helleman and R Stora (New York: North-Holland) 171 \new
Ozorio de Almeida A M 1988 {\sl Hamiltonian Systems, Chaos and Quantization}
  (Cambridge: Cambridge University Press).

\refis{HPage} Hawking S W 1985 Arrow of Time in Cosmology {\sl Phys. Rev.}
D {\bf 32} 2489-95\new
Page D 1985 Will Entropy Decrease if the Universe Recollapses?
{\sl Phys. Rev.} D {\bf 32} 2496-9\new
\dash\ 1991 The Arrow of Time, {\sl Proceedings of the First International
A. D. Sakharov Memorial Conference on Physics} M. Man'ko ed. (New York,
Nova Science).

\refis{Elskens} Prigogine I and Elskens Y 1987 {\sl Quantum Implications}
  ed. B J Hiley and F D Peat (London: Routledge) \new
Courbage M 1983 Intrinsic irreversibility of Kolmogorov dynamical systems
  {\sl Physica} {\bf122A} 459-82 \new
Calzetta E 1991 A necessary and sufficient condition for convergence to
  equilibrium in Kolmogorov systems {\sl J. Math. Phys.} {\bf32} 2903

\refis{Abbott} Abbott L F 1981 {\sl Nucl. Phys.} B {\bf 185},233.

\refis{mincoup}Belinsky V A, Grishchuk L P, Khalatnikov I M and Zel'dovich
  Ya. B. 1985 Inflationary Stages in Cosmological Models with a Scalar Field,
  {\sl Phys. Lett.} {\bf 155B}, 232-6\new
\dash\ 1985 (same  title) {\sl  Zh.  Eksp.  Teor.  Fiz.}{\bf 89} 346-60 (
  Engl. Trans. {\sl Sov. Phys. JETP} {\bf 62} 195-203)\new
\dash\ 1985 (same title) {Proceedings of the Third Seminar on Quantum
  Gravity} ed. M. A. Markov, V. A. Berezin and V. P. Frolov ( Singapore,
  World Scientific) 566-90\new
Gottlober S, Muller V and Starobinsky A 1991 Analysis of
  Inflation Driven by a Scalar Field and a Curvature Squared Term {\sl Phys.
  Rev.} D {\bf 43} 2510-20.

\refis{nonmin}Futamase T, Rothman T and Matzner R 1989
  Behavior of Chaotic Inflation in Anisotropic Cosmologies with
  Nonminimal Coupling {\sl Phys. Rev.} D {\bf 39} 405-11\new
Maeda K., Stein-Schabes J and Futamase T 1989 Inflation in a Renormalizable
  Cosmological Model and the Cosmic No Hair Conjecture {\sl Phys. Rev.} D
  {\bf 39} 2848-53\new
Amendola L, Litterio  M  and  Occhionero  F  1990  The  Phase Space View of
  Inflation (I) {\sl Int. J. Mod. Phys.} A {\bf 5} 3861-86\new
Demianski M 1991 Scalar Field, Nonminimal Coupling, and Cosmology {\sl
  Phys. Rev.} D {\bf 44} 3136-46\new
Demianski M, de Ritis R, Rubano C and Scudellaro P 1992 Scalar Fields and
Anisotropy in Cosmological Models {\sl Phys. Rev.} D {\bf 46} 1391-8.

\refis{Kiefer} Kiefer C 1989  Non  Minimally  Coupled Scalar Fields and the
Initial Value Problem in Quantum Gravity {\sl Phys. Lett.} {\bf 225 B}
227-32.

\refis{Parker} Parker L 1968 {\sl Phys. Rev. Lett.} {\bf 21}, 562\new
\dash\ 1969 {\sl Phys. Rev.}{\bf 183}, 1057.

\refis{Hartle} Hartle J B 1981 {\sl Phys. Rev.} D {\bf 23}, 2121\new
Calzetta E and Castagnino M 1984 Riemannian Approach and Cosmological
Singularity {\sl Phys. Rev. } D {\bf 29}, 1609-17.

\refis{me}Calzetta E 1991
Particle Creation, Inflation, and Cosmic Isotropy
{\sl Phys. Rev.} D {\bf 44}, 3043.

\refis{Diego}Calzetta E and Mazzitelli F 1990 Decoherence and Particle
Creation {\sl Phys. Rev.} D {\bf 42}, 4066-69.

\refis{Wolf}Wolf A, Swift J, Swinney  H and Vastano J 1985 Determining
Lyapunov Exponents from a Time Series {\sl Physica} {\bf 16D}, 285-317.

\refis{MTW} Misner C, Thorne K and Wheeler A 1972 {\sl Gravitation}
(San Francisco, Freeman).

\refis{Kuchar} Hartle J and Kucha\v r K 1984 The Role of Time in Path
Integral Formulations of Parametrized Theories, {\sl Quantum Theory
of Gravity}, ed. by S. Christensen (Bristol, Adam Hilger) pp. 315-26.

\refis{Courant} Courant R and Hilbert D 1953 {\sl Methods of Mathematical
Physics} (New York, Wiley) Vol I, p. 531.

\refis{Whittaker} Whittaker E and Watson G 1940 {\sl A Course of
Modern Analysis} (Cambridge, Cambridge University Press).

\refis{ReyPastor} Rey Pastor J, Pi Calleja P and Trejo C 1959
{\sl An\'alisis Matem\'atico} (Buenos Aires, Kapelusz), Vol III, pp.
176-88\new
McCracken D. and Dorn W 1964 {\sl Numerical Methods and FORTRAN Programming}
(New York, John Wiley).

\refis{Arnold} Arnold V I 1978 {\sl Mathematical Methods of Classical
  Mechanics} (Berlin: Springer-Verlag) \new
Arnold V I, Kozlov V V and Neishtadt A I 1988
  {\sl Mathematical Aspects of Classical and Celestial Mechanics}, Dynamical
  Systems III, Encyclopaedia of Mathematical Sciences
  (Heidelberg: Springer-Verlag)

\refis{Shields} Shields P 1974 {\sl The Theory of Bernoulli Shifts} (Chicago:
  University of Chicago Press)\new
Ornstein D 1974 {\sl Ergodic Theory, Randomness, and Dynamical Systems}
(New Haven, Yale University Press).

\endreferences
\vfill\eject

\centerline{\bf FIGURE CAPTIONS}

\bigskip

\noindent
{\bf Figure  1}  Evolution  of  the  scale  factor for a Robertson-Walker
cosmology in a weak chaotic situation
(${\phi}_i = 1$  and  ${\pi}_i = -1$).  After several oscilations there is
an unusual growth of the scale factor, this corresponds to an
``inflationary-like'' stage.

\noindent
{\bf Figure 2} Evolution  of the field $\phi$ corresponding to the inital
conditions of Fig.1.  It can be appreciated the oscillatory behavior with
a sudden change,  in frecuency and amplitude,
during the ``inflationary '' stages.

\noindent
{\bf Figure 3} Maximal Lyapunov exponent  (${\phi}_i  = 1$, ${\pi}_i =
-1$ and discarding a  transient  stage  near  $\eta  = 0$).  Its value is
clearly greater than zero.

\noindent
{\bf Figure 4} Maximal Lyapunov exponent  (${\phi}_i  = 10$, ${\pi}_i =
-10$ and discarding a transient stage near $\eta = 0$).

\noindent
{\bf Figure 5} Maximal Lyapunov exponent  (${\phi}_i  = 1$, ${\pi}_i =
-100$ and discarding a transient  stage  near  $\eta  =  0$).   It can be
appreciated  that  its  value  increase  with  the  increase  of  initial
condition, as predicted in section II.

\noindent
{\bf Figure 6} Projection of the phase space trajectory onto the $\phi$ ,
$a$ plane. In this situation
(${\phi}_i = 1$ and ${\pi}_i = -1$),  it behaves as a smoth Lissajous
curve.

\noindent
{\bf Figure 7} Same as Fig.6, (${\phi}_i =  1$  and  ${\pi}_i = -100$). The
trajectory is very irregular, as indicated by its high Lyapunov exponent.

\noindent
{\bf Figure 8} Same as Fig.6. This corresponds  to  the Lyapunov shown in
Fig.4  (${\phi}_i  = 10$ and ${\pi}_i = -10$).  See the increase of amplitude
in  the scale factor in the last two cases.

\noindent
{\bf Figure  9} Field at the Big Crunch against field at the Big Bang.
The  initial data are:  $a_i = 0.001$ (for computational
conveniences),  ${\pi}_i  =  -1$  and
${\phi}_i$ running from zero to one.  There is a simple  relation between
${\phi}_f$ and ${\phi}_i$.

\noindent
{\bf Figure 10} Same as Fig.9, with $a_i = 0.001$, ${\pi}_i  =  -10$  and
${\phi}_i$ running from zero to ten.  ``Islands'' of  stability may still
be seen, for example between (1.5 and 3), (3 and 4.5) and (5 and 6).

\noindent
{\bf Figure 11} Same as Fig.9, with $a_i = 0.1$, ${\pi}_i  =  -100$  and
${\phi}_i$ running from zero  to one hundred.  In this situation there is
no clear correlation between initial and final data.

\end